\documentclass[12pt]{article}
\usepackage{amssymb,amsmath,epsfig}

\begin{document}

\title{\bf Thermodynamics of Phantom Energy Accreting onto a Black
Hole in Einstein-Power-Maxwell Gravity}

\author{G. Abbas\thanks{Email:ghulamabbas@ciitsahiwal.edu.pk}
\\\\
Department of Mathematics, COMSATS Institute\\
of Information Technology, Sahiwal-67000, Pakistan.}
\date{}
\maketitle

In this paper, we investigate the phantom energy accretion onto $3D$
black hole formulated in Einstein-Power-Maxwell theory. We have
presented the conditions for critical accretion of phantom energy
onto black hole. Further, we discuss the thermodynamics of phantom
energy accreting onto black hole and found that first law of
thermodynamics is easily satisfied while second law and generalized
second law of thermodynamics remains invalid and conditionally
valid, respectively. The results for BTZ black hole can be recovered
by taking Maxwellian contribution equal to zero.\\


Recently, there has been a growing interest to study the black hole
(BH) solution in (2+1)-dimensions. These BH solutions have all the
typical properties that can be found in (3+1) or higher dimensions,
such as horizons, Hawking temperature and thermodynamics. The main
motivation to study the lower dimensional BH lies in the fact that
mathematical structure of such BHs is much simple as compared to
higher dimensional BHs. Therefore, one study analytically the the
properties of the BHs in lower dimension. On the other hand, the
nonlinear electrodynamic sources are valid tools for constructing
the regular BHs which have some extra properties like the validity
of first and zeroth laws of BH mechanics. On the basis of these
facts the study of $2+1$ dimensional BHs in nonlinear
electrodynamics has attracted a much attention \cite{1}-\cite{4}. In
any arbitrary $n$ dimensions Maxwell field has a power in the form
of $(F_{\mu\nu} F^{\mu\nu})^{n/4}$, which provides a traceless
Maxwell's energy momentum tensor \cite{5,6}. However, recently,
Gurtug et al. \cite{7}, have constructed a large class of most
general BH solutions in (2+1)-dimensions in Einstein-Power-Maxwell
(EPM) theory of gravity without imposing the traceless condition.

Several models for the explanation of dark energy (DE) were
suggested. These usually include quintessence, phantom, tachyon
field, holographic and brane-world models \cite{7a}-\cite{7d}. The
simplest form of DE is cosmological constant for which the equation
of state (EoS) parameter is $\omega=-1$, and hence pressure is
negative. The quintessence and phantom corresponds to $\omega>-1$
and $\omega<-1 $, respectively \cite{7e}. Phantom energy deviates
the null energy condition. The accelerating universe is dominated by
the phantom energy which diverges to approach the future
singularity.

In Newtonian theory, the problem of accretion of matter onto the
compact object was formulated by Bondi \cite{8}. Michel \cite{9}
formulated the relativistic model for the accretion of perfect fluid
onto the Schwarzschild BH. Babichev et al. \cite{10} examined that
phantom accretion onto a BH can decrease its mass, provided the back
reaction effects of accreting phantom energy on geometry of BH are
neglected. Jamil et al. \cite{11} discussed the critical accretion
on the Reissner-Nordstr$\ddot{o}$m (RN) BH. They determined a mass
to charge ratio beyond which a BH can be converted into a naked
singularity. This fact was also remarked by Babichev et al.
\cite{12}, by using the linear EoS and Chaplygin gas EoS for RN BH.
Madrid and Gonzalez \cite{13} showed that accreting phantom energy
onto Kerr BH can convert it into a naked singularity. Sharif and
Abbas \cite{14}-\cite{17e} discussed gravitational collapse and the
phantom energy accretion onto a class of BHs and found that CCH is
valid for phantom accretion onto a stringy charged BH. Jamil and
Akbar \cite{17a} have investigated the thermodynamics of phantom
energy accreting onto a BTZ BH. Recently, Abbas \cite {18} have
discussed the accretion of phantom energy onto a BH in
Ho$\check{r}$ava Lifshitz gravity. In this paper, we investigate the
thermodynamics of phantom energy accretion onto a static EPM BH by
using the procedure of Jamil and Akbar \cite{17a}.

We discuss the phantom energy accretion onto a (2+1) dimensional EPM
BH. From the study of phantom energy accretion, we can investigate
some interesting properties of EPM BH. The (2+1)$D$ action for EPM
theory with non-zero cosmological constant is given by \cite{7}
\begin{equation}\label{1}
S=\int dx^3\sqrt{-g}\left(\frac{1}{16\pi}(R-\frac{2}{3}\Lambda)-b_k
F^k\right),
\end{equation}
where $\verb"F"=F_{\mu\nu}F^{\mu\nu}$ is the Maxwell invariant and
$k$ is an arbitrary rational number. Taking $ k=\frac{2}{3}$, the
metric for (2+1)$D$ is \cite{19}

\begin{equation}\label{1}
ds^2=A(r)dt^2- \frac{dr^2}{A(r)}-r^2d{\theta}^2,
\end{equation}
where $A(r)=-M+\frac{r^2}{l^2}+\frac{4\pi Q^{4/3}}{3r^2}$, where
$l^2=-\frac{1}{\Lambda}>0$, as $\Lambda<0$. The horizons of the BH
can be determining by the zeros of laps functions, i.e., $A(r)=0$,
this leads to
\begin{equation}
r_{\pm}=\sqrt{\frac{3Ml^2\pm\sqrt{9M^2l^4-48\pi l^2Q^{4/3}}}{6}}
\end{equation}
where $r_{-}$ and $r_{+}$ represent inner and outer horizons of BH,
respectively. Both the horizons can exist only if
$\frac{9M^2l^2}{48\pi Q^{4/3}}\geq\frac{48}{9}$. The regularity of
this BH can be seen in the regions $r_{+}<r<\infty,~r_{-}<r<r_{+}$
and $0<r<r_{+}$.

The energy momentum tensor of phantom energy in the form of perfect
fluid given by
\begin{equation}\label{7}
{T_{{\mu}{\nu}}={({\rho}+p)}u_{\mu}u_{\nu}-pg_{\mu\nu}},
\end{equation}
where $\rho$ and $p$ are the density and pressure of the accreting
matter and $u^\mu=(u^0,u^1,0)$ is the velocity vector of the
infalling fluid. It is mentioned here that $u^\mu$ satisfies the
normalization condition, i.e., $u^\mu u_\mu =1$.

For the phantom energy accretion onto a EPM BH, we shall formulate
two equations of motion for the fluid in the vicinity of BH. The
energy conservation equation $T^{t\mu}_{;\mu}=0$ is given by
\begin{equation}
\label{8} ru(\rho+p)\left(-M+\frac{r^2}{l^2}+\frac{4\pi
Q^{4/3}}{3r^2}+u^2\right)^{\frac{1}{2}}=A_0,
\end{equation}
where $A_0$ is an integration constant and $u^{1}=u<0$ as the matter
flow is directed inward from outer source onto EPM BH. Also, by
projecting the energy-momentum conservation law on velocity, we get
the energy flux flowing onto the BH. Thus Eq.(\ref{7}) leads to
\begin{equation}\label{9}
ru\exp
\left[\int^\rho_{\rho_\infty}\frac{d\rho'}{\rho'+p(\rho')}\right]=-A_1,
\end{equation}
where $A_1>0$ is another integration constant. Also, ${\rho}$ and
${\rho_\infty}$ are densities of the phantom energy at $r=finite$
and $r\rightarrow\infty$. From Eqs.(\ref{8}) and (\ref{9}), we
obtain
\begin{equation}\label{10}
(\rho+p)\left(-M+\frac{r^2}{l^2}+\frac{4\pi
Q^{4/3}}{3r^2}+u^2\right)^{\frac{1}{2}}
\exp\left[-\int^\rho_{\rho_\infty}\frac{d\rho'}{\rho'+p(\rho')}\right]=A_2,
\end{equation}
where $A_2=-\frac{A_0}{A_1}=\rho_\infty +p(\rho_\infty)$.

Now we analyze the critical points (such points at which flow speed
is equal to the speed of sound) during the accretion of phantom
energy. The phantom energy falls onto BH with increasing velocity
along the particle trajectories. For the discussion of critical
accretion points, we follow the procedure of Michel \cite{9}. The
conservation of mass flux
\begin{equation}\label{13}
\rho u r=E_0,
\end{equation}
where $E_0$ is the constant of integration. Dividing and squaring
Eqs.(\ref{8}) and (\ref{13}), we get
\begin{equation}\label{14}
\left(\frac{\rho +p}{\rho}\right)^2
\left(-M+\frac{r^2}{l^2}+\frac{4\pi Q^{4/3}}{3r^2}+u^2\right)=D_1,
\end{equation}
where $E_1=(\frac{A_0}{E_0})^2$ is a positive constant.
Differentiating Eqs.(\ref{13}) and (\ref{14}) and eliminating
$d\rho$, we get
\begin{equation}\label{15}
\frac{dr}{r}\left[V^2-\frac{(\frac{r^2}{l^2}-\frac{4\pi
Q^{4/3}}{3r^2})} {A(r)+u^2}\right]+\frac{du}{u}
\left[V^2-\frac{u^2}{A(r)+u^2}\right]=0.
\end{equation}
where $V^2=\frac{d\ln(\rho+p)}{d\ln\rho}-1$ and $A(r)$ is the lapse
function as defined after Eq.(\ref{1}).

This equation shows that critical points are located at such values
of $r$ (say at $r_c$) where both the square brackets vanish. Thus
\begin{equation}\label{16}
{V_c}^2=\frac{(\frac{r^2}{l^2}-\frac{4\pi Q^{4/3}}{3r^2})}
{A(r)+u^2} ,\quad {V_c}^2=\frac{u^2}{A(r)+{u_c}^2}.
\end{equation}
From above equations, we get

${u_c}^2>(\frac{r^2}{l^2}-\frac{4\pi Q^{4/3}}{3r^2})$ and
${V_c}^2>\frac{{u_c}^2}{A(r)+{u_c}^2}$. We mentioned that the
physically acceptable solutions of the above equations are obtained
if ${u_c}^2>0$ and ${V_c}^2>0$ leading to
\begin{eqnarray}\label{18}
3{r_c}^4-4\pi Q^{4/3}l^2>0,\quad -M+\frac{2r^2}{l^2}>0.
\end{eqnarray}
From the first equation, we get
${r_c}^2=\frac{2}{\sqrt{3}}\sqrt{\pi}Q^{2/\sqrt{3}}l$, using this in
second equation of Eq.(\ref{18}), we have
$\frac{M^2l^2}{Q^{4/3}}\geq\frac{16l^2}{9}$, for this ratio horizons
disappear, hence we have naked singularity at $r=0$.

Here, we discuss the thermodynamics of phantom energy that crosses
the event horizons of EPM BH. For this purpose, we follow \cite{17a}
to write EPM BH metric in the following form
\begin{equation}\label{G1}
ds^2=h_{\alpha \beta}dx^{\alpha} dx^{\beta}-r^2d\theta^2,\quad
\alpha,\beta=0,1,
\end{equation}
where $h_{\alpha \beta}=diag(A(r),\frac{1}{A(r)})$, is a
2-dimensional metric. From the normalized condition of velocities,
i.e., $u^\mu u_\mu =1$, we can get the relation
\begin{equation}\label{G2}
u^0=A(r)^{-1}\sqrt{A(r)+u^2},\quad u_0=\sqrt{A(r)+u^2}.
\end{equation}
The components of stress energy tensor are
\begin{equation}\label{G12}
T^{00}=A(r)^{-1}[(\rho+p)\frac{A(r)+u^2}{A(r)}-p]
\end{equation}
\begin{equation}\label{G20}
T^{11}=[(\rho+p)u^2+A(r)p].
\end{equation}
These components will be used to calculate the work density function
defined by $W=\frac{1}{2}T^{\alpha \beta}h_{\alpha \beta}$, for the
metric Eq.(\ref{G1}), it turn as
\begin{equation}\label{G2}
 W=\frac{1}{2}(\rho-p).
\end{equation}
Now the energy supply vector is
\begin{equation}\label{G2}
 \Psi_{\beta}={T^{\alpha}}_{\beta}\partial_{\alpha} r +W \partial_{\beta} r.
\end{equation}
The components of the energy supply vector are
\begin{equation}\label{G26}
\Psi_0=-u(\rho+p)\sqrt{A(r)+u^2}
\Psi_1=(\rho+p)\left(\frac{1}{2}+\frac{u^2}{A(r)}\right).
\end{equation}
The change of energy across the apparent horizon is defined as
$-dE=-A\Psi$, where $\Psi=\Psi_0 dt+\Psi _1dr$. The energy crossing
the event horizon of the BH in time interval dt is given by
\begin{equation}\label{G2}
 dE=2\pi r_{e}u^2(\rho+p)dt.
\end{equation}
where $r_e=r_+$. Assuming $E=M$, the comparison of Eqs.(\ref{10})
and (\ref{G2}), we get $A_2=u^2r_e$. The entropy of the EPM BH is
given by
\begin{equation}\label{G3}
 S_h=2\pi r_{e}.
\end{equation}
It can be verified that thermal quantities i.e., change of accreting
phantom energy $dE$, horizon entropy $S_h$ and horizon temperature
obeys the first law of thermodynamics, $dE=T_hdS_h$. Differentiating
the last equation with respect to $t$, using equation (\ref{10}), we
get
\begin{equation}\label{G4}
 \dot{S}_h=\pi u^2(\rho+p)l^2\left(\frac{1+\frac{1}{\sqrt{9M^2l^4-48\pi l^2Q^{4/3}}}}{12}\right).
\end{equation}

As all the parameters are positive in above expression except
$(\rho+p)<0$ (accreting matter under discussion is phantom energy
foe which weak energy condition is violated). Hence phantom energy
accreting onto EPM BH does not obey the second law of
thermodynamics.

Now we check the validity of second law of thermodynamics. It is
given by

\begin{equation}\label{G4}
 \dot{S}_{total}=\dot{S}_{h}+\dot{S}_{ph}\geqslant0.
\end{equation}
This can be explained as follows: The sum time rate of change of
entropies BH horizon and phantom energy should be positive. We that
event horizon of EPM BH behaves as the boundary of thermal system
and total energy in the interior of the horizon is the is the mass
of the EPM BH. Further, following \cite{17a}, we consider
\begin{equation}\label{G6}
T=T_{h}=T_{ph},
\end{equation}
 (horizon temperature is in equilibrium with
temperature of phantom), where $T_{ph}$ and $T_{h}$ are temperature
of phantom energy and horizons, respectively. It is well known that
Einstein field equations satisfy the first law of thermodynamics,
\begin{equation}\label{G5}
T_{h}{dS_h}=pdA+dE
\end{equation}
at the event horizon. We assume that above equation holds well for
phantom energy accreting onto the BH, i.e.,
\begin{equation}\label{G7}
T_{ph}{dS_{ph}}=pdA+dE
\end{equation}
The temperature at event horizon of EPM BH is \cite{19}
\begin{equation}\label{G7}
T_{h}=\frac{r_e}{2\pi l^2}-\frac{2Q^{4/3}}{3r^3_e}.
\end{equation}
From Eqs.(\ref{G4}) and (\ref{G6}), we get

\begin{equation}\label{G666}
 T \dot{S}_{total}=T(\dot{S}_{h}+\dot{S}_{ph})=\pi u^2l^2(\rho+p)\left(2+\frac{\pi p}{6r_e}
 \left(1+\frac{3lM}{\sqrt{9l^4M^2-48\pi Q^{4/3}}}\right)\right).
\end{equation}
Since for phantom energy $(\rho+p)<0$,  and all the remaining
parameters are positive, so generalized second law of thermodynamics
will be valid in this case if
$p<-\frac{12r_e}{\pi\left(1+\frac{3lM}{\sqrt{9l^4M^2-48\pi
Q^{4/3}}}\right)}$. Since $p<0$ for phantom energy, so validity of
generalized second law of thermodynamics provides the upper bound on
pressure of phantom energy.

In summary, following the method of Jamil and Akbar \cite{17a}, we
discuss the accretion and critical accretion phenomena of phantom
energy onto EPM BH. The detailed analysis of mass conservation and
energy flux equations implies that the mass of BH decreases due to
infall of phantom energy and consequently EPM gravity BH is
converted into naked singularity. As for
$\frac{M^2l^2}{Q^{4/3}}\geq\frac{16l^2}{9}$, horizons become
imaginary and consequently disappear and singularity becomes naked
at $r=0$.

There exists two horizons, event horizon at $r=r_+$ and Cauchy
horizon at $r=r_-$) of EPM gravity BH for $\frac{9M^2l^2}{48\pi
Q^{4/3}}\geq\frac{48}{9}$, otherwise there would occur a naked
singularity. We have found that the existence of Cauchy horizon
requires $Q\neq0$ like Reissner-Nordstr$\ddot{o}$m BH. If we take
$Q=0$ then there will be a unique horizon at $r=\sqrt{M}~l$, which
is the horizon of BTZ BH. The critical accretion analysis implies
that corresponding to two horizons there exists a two values of $r$,
which provide the physically acceptable critical point for
$\frac{M^2l^2}{Q^{4/3}}\geq\frac{16l^2}{9}$. For this ratio the
horizons of the EPM BH disappear and we have naked singularity hence
Cosmic Censorship is invalid in this case.

Also, we have discussed in detail the thermodynamical behavior of
phantom energy accreting onto EPM gravity BH. It has been found that
first law of thermodynamics holds in general for the assumed matter
accreting onto EPM gravity BH while second law of thermodynamics is
invalid in this case. By assuming that event horizon of BH behaves
as the boundary thermal system and temperature of horizon is in
equilibrium with the temperature of the phantom at the horizon, we
have derived the GSL of thermodynamics. Under the above mentioned
constraints, we have proved that GSL holds if the pressure of
phantom energy has a bound
$p<-\frac{12r_e}{\pi\left(1+\frac{3lM}{\sqrt{9l^4M^2-48\pi
Q^{4/3}}}\right)}$ on the parameters of EPM gravity BH. The results
for BTZ BH can be recovered by taking $Q=0$.

\end{document}